\PassOptionsToPackage{table,xcdraw}{xcolor}
\documentclass[manuscript]{acmart}

\usepackage{tabularx}
\usepackage{multirow}
\usepackage{subcaption}

\copyrightyear{2024}
\acmYear{2024}
\setcopyright{rightsretained}
\acmConference[IUI Companion '24]{29th International Conference on Intelligent User Interfaces - Companion}{March 18--21, 2024}{Greenville, SC, USA}
\acmBooktitle{29th International Conference on Intelligent User Interfaces - Companion (IUI Companion '24), March 18--21, 2024, Greenville, SC, USA}
\acmDOI{10.1145/3640544.3645258}
\acmISBN{979-8-4007-0509-0/24/03}

\begin{document}

\title{Speech as Interactive Design Material (SIDM)}
\subtitle{How to design and evaluate task-tailored synthetic voices?}

\author{Mateusz Dubiel}
\orcid{0000-0001-8250-3370}
\affiliation{%
  \institution{University of Luxembourg}
  \country{Luxembourg}}
\email{mateusz.dubiel@uni.lu}

\author{Matthew Aylett}
\orcid{0000-0001-7057-0525}
\affiliation{%
  \institution{Heriot Watt University and CereProc}
  \country{UK}}
\email{m.aylett@hw.ac.uk}

\author{Anuschka Schmitt}
\orcid{0000-0003-1101-6529}
\affiliation{%
  \institution{University of St.Gallen}
  \country{Switzerland}}
\email{anuschka.schmitt@unisg.ch}

\author{Zilin Ma}
\orcid{0000-0002-7259-9353}
\affiliation{%
  \institution{Harvard University}
  \country{USA}}
\email{zilinma@g.harvard.edu}

\author{Gary Hsieh}
\orcid{0000-0002-9460-2568}
\affiliation{%
  \institution{University of Washington}
  \country{USA}}
\email{garyhs@uw.edu}

\author{Thiemo Wambsganss}
\orcid{0000-0002-7440-9357}
\affiliation{%
  \institution{Bern University of Applied Sciences}
  \country{Switzerland}}
\email{thiemo.wambsganss@bfh.ch}

\renewcommand{\shortauthors}{Dubiel et al.}

\begin{abstract} 
The aim of this workshop is two-fold. First, it aims to establish a research community focused on design and evaluation of synthetic speech (TTS) interfaces that are tailored not only to goal oriented tasks (e.g., food ordering, online shopping) but also personal growth and resilience promoting applications (e.g., coaching, mindful reflection, and tutoring). Second, through discussion and collaborative efforts, to establish a set of practices and standards that will help to improve ecological validity of TTS evaluation. In particular, the workshop will explore the topics such as: interaction design of voice-based conversational interfaces; the interplay between prosodic aspects (e.g., pitch variance, loudness, jitter) of TTS and its impact on voice perception. This workshop will serve as a platform on which to build a community that is better equipped to tackle the dynamic field of interactive TTS interfaces, which remains understudied, yet increasingly pertinent to everyday lives of users.
\end{abstract}



\maketitle

\section{Theme and Goals}
The theme of the workshop is ``Speech as a Design Material for Interaction: How to design and evaluate task-tailored synthetic voices?'' Increasingly, AI-based artifacts are embedded in interfaces that interact with the user in a dialogue-based manner, via text and voice. Indeed, voice-based user interaction with AI-based artifacts can be viewed as an emerging and essential facet of HCI~\cite{clark2019state}.  The growing popularity and increased usage of interfaces that feature synthetic speech could be partly attributed to their ever-improving
natural language processing capabilities. Recent developments in Deep Learning have contributed to a rapid improvement of the quality of synthetic voices in terms of intelligibility and
naturalness, making them almost indistinguishable from human speech~\cite{gibiansky2017deep}. Conversational Agents such as Amazon Alexa and Google Home are already being used for habitual activities that can directly affect user finances (e.g., takeaway-ordering or purchasing products online) \cite{dubiel2022conversational}. However, despite constant technological developments in speech technology and commercial deployment, there is still a paucity of research examining the voice design and its implications for the user \cite{schmitt2021voice}. Consequently, there is a lack of consistency in how synthetic voices are developed and evaluated through interactive user studies (both online and in the lab). 

In order to address this problem, the purpose of this workshop is to bring interdisciplinary experts from audio engineering, speech perception, UX design, and other relevant fields together to improve the design practices of user interfaces that feature synthetic speech and increase the ecological validity of their evaluation. Specifically, during the workshop, through interactive demonstrations, group discussions and prototyping activities, we will investigate the following questions:

\begin{itemize}
    \item \textit{How to ensure a comprehensive and comparable design, development, and evaluation of synthetic voices for user interfaces?}
    \item \textit{How to effectively tailor voices to different domains of application?}
    \item \textit{Which auditory cues are crucial to investigate, for what tasks, and in which contexts?}
\end{itemize}

We encourage collaboration between academia and industry and welcome participation from anyone interested in synthetic speech design, development, and evaluation.  By bringing the interdisciplinary community together, our goal is to provide insights on how to improve engineering of speech interfaces that are tailored to specific tasks, and ensure higher ecological validity of their evaluation.

\section{Background and Motivation}
Voice design, among other fields, is studied in human-agent interaction, phonetics and socio-linguistics, and has considered how auditory cues such as pitch variance or speaking rate and they modifications affect user perception and behaviour (e.g. ,\cite{chidambaram2012designing, dubiel2020persuasive,elkins2013sound,mcaleer2014you, belin2017sound}). In this workshop our focus is on such auditory cues that affect how the synthetic voice sounds, rather than stylistic and lexical aspects of the message that it conveys. 

Conceptual frameworks and voice design recommendations exhibit several gaps. As posited by Cambre and Kulkarni \cite{cambre2019one}, existing frameworks and reviews do not provide a mapping of CA-relevant voice cues against user outcomes and their evaluation through relevant experiments. To overcome this gap we propose a workshop that will bring audio engineers, UX designers, and speech perception experts together to discuss how to design and implement interactive speech evaluation studies test the impact of specific acoustic cues and increase ecological validity of such evaluations. For instance, by creating CAs with synthetic voices tailored to specific task and evaluate them in the context of their intended usage.

\section{Organisers}
\noindent \textbf{Mateusz Dubiel} is a Research Associate in the Department of Computer Science at the University of Luxembourg, where he works on development and evaluation of conversational agents. Specifically, his current research focuses on assessment of cognitive and usability implications of interfaces that feature speech, and exploration of their potential to inspire positive behavioural change in users. 

\noindent \textbf{Matthew Aylett} is an Associate Professor at Heriot Watt Edinburgh and a co-founder and Chief Scientific Officer of CereProc. His work focuses on human/robot interaction and conversational interaction. 

\noindent \textbf{Anuschka Schmitt} is a research associate and Ph.D. candidate at the Institute of Information Management at the University of St.Gallen (HSG), Switzerland and currently a visiting researcher at Harvard University's School of Engineering and Applied Sciences. Her research focuses on user trust and decision-making in human-computer interaction, as well as perceptual and behavioural implications of conversational AI. 

\noindent \textbf{Zilin Ma} is a Ph.D. student in the Department of Computer Science at Harvard University. He works in the bias issues associated with voice assistants. His current research project looks at how the perception of accent affects one's trust on voice assistants, and how it affects decisionmaking alone with AI. His work has implications in how we design equitable voice interface to benefit broader populations.

\noindent \textbf{Gary Hsieh} is an Associate Professor in the Department of Human-Centred Design and Engineering (HCDE) and an Adjunct Associate Professor in Paul G. Allen School of Computer Science and Engineering at the University of Washington. He specialises in designing and developing technologies
that encourage people to communicate and interact in ways that are self- and welfare-improving. His research
extends and builds on theories from many research fields, including psychology, communication, and economics.

\noindent \textbf{Thiemo Wambsganss} is an Assistant Professor at Bern University of Applied Sciences. His work aims to leverage methods from Natural Language Processing and Machine Learning to provide users, in particular students, with intelligent writing feedback anytime and anywhere they want. In this vein, he strives to understand how humans perceive, interact, and learn with intelligent tools, such as with conversational voice assistance for smart data collection.

\section{Schedule and description of activities planned}
The tentative schedule and workshop activities are presented in \autoref{tab:activities}. Since the purpose of our workshop is to  synthesise insights from experts from different communities and translate them to practical design and evaluation recommendations, we will ensure that ample time is dedicated to prototyping activities, discussions and interactions between the participants. 

\begin{table}[h!]
\caption{Tentative Workshop Plan}
\begin{tabular}{lll}
\toprule
\multicolumn{1}{c}{\begin{tabular}[c]{@{}c@{}}\textbf{Time  Slot (duration)} \end{tabular}} & \multicolumn{1}{c}{\begin{tabular}[c]{@{}c@{}}\textbf{Activity}\end{tabular}} &  \\ \midrule
09:00-09:15~(15mins)  & Welcome                         &  \\
09:15-10:15~(60mins)  & Keynote by Prof. Benjamin Cowan               &  \\
10:15-10:30~(15mins) & Coffee Break                       &  \\
10:30-11:00~(30mins) & Expressive TTS Demonstration    &  \\
11:00-12:00~(60mins) & Group Prototyping Session (I)    &  \\
12:00-13:00~(60mins) & Lunch Break                     &  \\
13:00-14:15~(45mins)  &  Group Prototyping Session (II)             &  \\
14:15-14:30~(15mins) & Coffee Break                      &  \\
14:30-16:00~(90mins)  & Group Reporting and Discussion                 &  \\
16:00-16:45~(45mins) & Closing Remarks and Future Plans  &  \\ \bottomrule
\end{tabular}
\label{tab:activities}
\end{table}

In order to inspire participants and spur discussions, the workshop will begin with a keynote delivered by Prof. Benjamin Cowan from University College Dublin\footnote{https://people.ucd.ie/benjamin.cowan} who will discuss the challenges of designing and building expressive speech interfaces and their potential for behavioural change. The morning part of the workshop will also include a demo session, where participants will be presented with samples of expressive synthetic voices developed with the state-of-the-art neural TTS software. The demonstration will be followed by the group prototyping session (part I), were participants will work together to design their own synthetic voice for a specific application. To foster interdisciplinary collaboration, each group will include participants from different backgrounds. In order to facilitate the voice development process and assist participants with any questions, each group will be moderated by at least one workshop organiser.

In the afternoon, after lunch break, group prototyping activities will continue. The participants will have 45 minutes to finalise their speech samples and prepare short presentations motivating the design process and proposing speech evaluation criteria. Then after a coffee break, each group will give a short presentation showcasing their synthetic voices and explaining their fit to the proposed interactive scenario (60 minutes), followed by 30 minutes of brainstorming activity and summarising the insights. The exact format and duration of group reporting and discussion will be based on final number of the workshop attendees (max. 20 people). 

\section{Anticipated Outcomes}
We anticipate that our workshop will provide both theoretical and practical outcomes. As for theory, the workshop will: (1) bring a better, cross-disciplinary understanding on how synthetic voices should be developed to provide the best-fit for the designated task and ensure best user experience, and (2) help to inform a set of standardised evaluation criteria to ensure that systems that feature such voices are evaluated in a more ecologically valid way.

In terms of practical outcomes, `Speech as a Design Material for Interaction' workshop will  lead to formation of new research community comprised of academics, industry experts and members of the public that will promote research on the interactive, task-based evaluation of synthetic speech and continuously identify and tackle new challenges as they arise. Ultimately,  we anticipate that concerted efforts of our newly formed community will lead to higher robustness and standardisation of speech evaluation experiments design. 
In summary, the workshop will lead to:

\begin{enumerate}
    \item Fostering a closer collaboration between the IUI community members and speech perception researchers.
    \item Synthesising views from different areas of expertise (i.e., audio engineering, UX desing and speech perception).
    \item Better design and development practice, and  improved ecological validity of interactive speech evaluation studies.
\end{enumerate}

\section{Format and plans for asynchronous engagements}
While we hope that most of participants will join the workshop in person, we will also provide a synchronous online access. To facilitate remote attendance, we will use Zoom for synchronised talks and Slack for online discussions and asynchronous question answering. We will work together with the technical team at IUI 2024 to utilise provided streaming channels to reduce jumps across the platforms.

\section{Results Dissemination beyond IUI 2024}
The workshop outcomes will be presented as a report. To reach a broader audience, the workshop will be recorded
and posted on the workshop website. Additionally, the outcomes of the workshop will be published as Medium posts. The organisers will use social media platforms (Mastodon and X (formerly Twitter)) to disseminate these materials.

\bibliographystyle{ACM-Reference-Format}

\end{document}